\documentclass[twocolumn,prl,aps,superscriptaddress]{revtex4}
\usepackage{amsmath}
\usepackage{graphicx}
\usepackage{amssymb}

\usepackage{amsmath}
\usepackage{graphicx}
\usepackage{amssymb}

\begin{document}

\title{A Josephson Junction Microscope for Low-frequency Fluctuators}

\author{L. Tian}
\email{ltian@stanford.edu}
\affiliation{Department of Applied Physics and E. L. Ginzton Laboratory, Stanford University, Stanford, CA 94305}
\author{R. W. Simmonds}
\email{simmonds@boulder.nist.gov}
\affiliation{National Institute of Standards and Technology, 325 Broadway,  Boulder, Colorado 80305-3328, USA}

\date{\today{}}

\begin{abstract}
The high-Q harmonic oscillator mode of a Josephson junction can be used as a novel probe of spurious two-level systems (TLSs) inside the amorphous oxide tunnel barriers of the junction.  In particular, we show that  spectroscopic transmission measurements of the junction resonator mode can reveal how the coupling magnitude between the junction and the TLSs varies with an external magnetic field applied in the plane of the
tunnel barrier.  The proposed experiments offer the possibility of clearly resolving the underlying coupling mechanism for these spurious TLSs, an important decoherence source limiting the quality of superconducting quantum devices.
\end{abstract}
\maketitle

Superconducting quantum circuits have been intensively tested in various regimes in the past few years, from superconducting qubits demonstrating long coherence times, to superconducting transmission line cavities coherently coupled to a Single Cooper Pair box \cite{MakhlinRMP2001,OliverScience2005,ChiorescuNature2004,KochPRL2006,WallraffNature2004,SteffenPRL2006}. Such circuits are extremely sensitive to very small quanta and defect states, and hence have the ability to detect individual microwave photons, charged quasiparticles, as well as  spurious TLSs within or near Josephson junction tunnel barriers \cite{SchusterNature2007,NaamanPRL2006,MartinisPRL2005,SimmondsPRL2004,PlourdePRB2005}.   In recent experiments \cite{MartinisPRL2005,SimmondsPRL2004}, TLSs were identified through spectroscopic measurements of a superconducting phase qubit appearing as `gaps" or ``splittings" in the energy spectrum.  

The TLS defects can be an unwanted source of decoherence for superconducting quantum bits.  The low-frequency noise, which has been shown to be a serious
source of decoherence for superconducting qubits \cite{AstafievPRL2006,MartinisPRL2005,SimmondsPRL2004,HarlingenPRB2004,WellstoodAPL2004},  is very probably induced by such amorphous fluctuators inside or near Josephson junctions \cite{DuttaRMP1981,WeissmanRMP1988}. Understanding the origin of these spurious TLSs, their coherent quantum behavior, and their connection to ubiquitous 1/f noise is hence a challenge that will be crucial  to the future of superconducting quantum devices.  The behavior of a distribution of these TLSs were
studied theoretically in \cite{MartinPRL2005,ShnirmanPRL2005,KochCondMat0702025}.
Recently, it was proposed that TLSs can be viewed as qubits themselves \cite{ZagoskinPRL2006}, given their relatively long coherence times. However, the microscopic origin and the coupling mechanism between the TLSs and the junction remains unresolved.  Generally considered to be connected to the amorphous nature of the tunnel barrier \cite{SOhPRB2006}, movement of unrestrained atoms or charges may lead to a number of possible coupling mechanisms. As originally proposed in Ref. \cite{SimmondsPRL2004}, fluctuations of the TLS could lead to variations of the junction
critical current. Another possibility, requires that the TLSs have fluctuating dipole moments which couple to the electric field found within the junction tunnel barrier \cite{MartinisPRL2005}. 

Here, we present a scheme that can resolve a variety of properties of the TLSs and  distinguish between these two suggested coupling mechanisms through the use of an applied magnetic field.  Consider a Josephson junction resonator operating as a high-Q, nonlinear cavity mode \cite{OsbornASC2006}, coupling to a TLS through its canonical phase (or momentum) operator. This forms a cavity QED system with the junction resonator mode as the cavity and the TLS as the atom \cite{RaimondRMP2001, HoodScience2000}.  The junction resonator acts as a microscope for studying  the behavior of the TLS.  We will show that microwave transmission in the junction resonator carries spectroscopic, interaction (coupling), and spatial information of the TLS.    In particular, the junction resonator can resolve the coupling mechanism between the TLS and the junction.   When the TLS couples to the junction through the junction \emph{critical current}, the magnitude of the coupling will be strongly modulated by changing the strength of the magnetic field oriented along the plane of the tunnel barrier. On the contrary, if the TLS can only couple through the junction's \emph{electric field}, magnetic fields will have no effect on the magnitude of this coupling.  Changes in the coupling magnitude can be observed by measuring the microwave transmission through the junction resonator.  To demonstrate this quantitatively, we calculate the transmission  and its noise spectrum in the ``bad cavity'' limit \cite{WangVyasPRA1995}, where the dissipation of the junction resonator is much faster than that of the TLS. Our calculations show that the resonances in transmission and the noise spectrum strongly depend on the coupling strength for chosen TLS. Meanwhile, the energy distribution, dynamic, dissipative, as well as spatial  properties of the TLSs can be obtained from the measured transmission as well.

\begin{figure}
\includegraphics[%
  bb=90bp 510bp 480bp 654bp,
  clip,
  width=7.5cm]{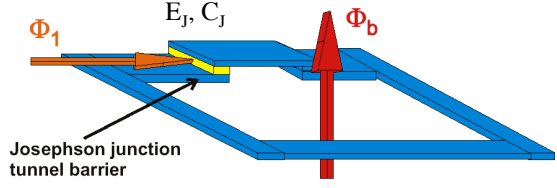}
\caption{Circuit. RF-SQUID loop with in-plane magnetic flux $\Phi_1=\frac{\hbar}{2e}\varphi_1$ and loop flux: $\Phi_b$. The Josephson junction has energy $E_J$ and capacitance $C_J$. }\label{fig1}
\end{figure}

\emph{Magnetic field modulation of the coupling.} %
The circuit is shown in Fig. \ref{fig1}, where a Josephson junction with the energy $E_{J1}$  together with a superconducting  loop  forms an RF SQUID enclosing a bias flux $\Phi_b$. Magnetic field applied along the plane of the tunnel barrier, inside the junction
\cite{OrlandoBook2001}, results in a flux of $\Phi_1=\frac{\hbar}{2e}\varphi_{1}$. At $\Phi_b=0$, the effective Josephson energy of the circuit can be written as
$-E_{J}\cos(\delta + \varphi_1/2)$ with\[
E_{J}=E_{J1}\frac{\sin\varphi_{1}/2}{\varphi_{1}/2} \]
modified by the magnetic field and $\delta$ being the  phase difference across the junction.  Given a total capacitance $C_{J}$, the junction behaves as an harmonic oscillator for the phase variable $\delta$ with a frequency $\omega_{c}=\sqrt{4e^{2}E_{J}/\hbar^{2}C_{J}}$.    Consider a TLS inside the junction with an energy  $\hbar\omega_{a}$ close to $\hbar\omega_{c}$. When
the TLS couples with the critical current of the junction, the coupling can be derived as\[
-E_{J1}\int_{0}^{L}dx\cos(\delta+\varphi_{1}\frac{x}{L})\vec{j}_{d}\cdot\vec{\sigma}f(x-r_{d})\]  where  $L$ is the length of the junction along the direction perpendicular to $\Phi_1$ in Fig. \ref{fig1}, $\vec{j}_{d}$ describes  the polarization and magnitude of the coupling, $\vec{\sigma}=(\sigma_{x},\sigma_{y},\sigma_{z})$ are the Pauli matrices of the TLS, and $f(x-r_{d})$ describes the spatial profile of the TLS centered at $r_{d}$ \cite{DuttaRMP1981,WeissmanRMP1988}. For simplicity, we assume $f(x-r_{d})=\delta(x-r_{d})$ and $\vec{j}_{d}=(j_{x},0,0)$. The coupling becomes $-E_{J1}j_{x}\cos(\delta+\varphi_{1}r_{d}/L)\sigma_{x}$. To the lowest order of the shifted phase variable $(\delta+\varphi_{1}/2)$, the coupling can be written as $H_{c}=g_{d}\sigma_{x}(\hat{a}+\hat{a}^{\dagger})$ with
\begin{equation}
g_{d}=E_{J1}j_{x}\sqrt{\frac{2e^{2}}{C_{J}\hbar\omega_{c}}}\sin\varphi_{1}(\frac{r_{d}}{L}-\frac{1}{2})\label{eq:gd}\end{equation}
and $\hat{a}$ and $\hat{a}^{\dagger}$ being the annihilation and creation operators of the phase variable. The coupling hence oscillates with and is strongly affected by the applied flux $\Phi_{1}$.

In contrast, when the TLS couples to the dielectric field within the Josephson junction, the coupling would be $-\frac{2e^{2}d_{0}}{C_{J}h_{0}}\frac{\hat{p}_{\delta}}{\hbar}$, where
$\hat{p}_{\delta}$ is the conjugate of the phase variable $\delta$, $d_{0}$ is the size of the dielectric dipole, and $h_{0}$ is the thickness of the junction tunnel barrier. The resulting coupling can be written as $H_{c}=-ig_{c}\sigma_{x}(\hat{a}-\hat{a}^{\dagger})$
with\begin{equation}
g_{c}=\frac{d_{0}}{h_{0}}\sqrt{\frac{e^{2}\hbar\omega_{c}}{2C}}\label{eq:gc}\end{equation}
not depending on the applied magnetic flux $\Phi_{1}$.  Therefore, probing the dependence of the coupling on the magnetic flux will clearly determine which physical mechanism is more visible even if \emph{both} effects are present. Note that  the applied  magnetic field $\Phi_1$ shifts the frequency of the junction resonator mode which can be tuned by controlling the loop flux $\Phi_b$  (see below).

The driving on the junction resonator can be obtained by capacitively coupling the junction to a microwave source of $2\epsilon\cos\omega_{d}t(\hat{a}+\hat{a}^{\dagger})$ with frequency $\omega_{d}$ and amplitude $\epsilon$. In the rotating frame, the total Hamiltonian for the coupled TLS and junction resonator system is\begin{equation}
H_{t}=\Delta_{c}\hat{a}^{\dagger}\hat{a}+\frac{\Delta_{a}}{2}\sigma_{z}+g(\varphi_{1})(\sigma_{+}\hat{a}+\sigma_{-}\hat{a}^{\dagger})+\epsilon(\hat{a}+\hat{a}^{\dagger})\label{eq:Ht}\end{equation}
with $\Delta_{c}=\omega_{c}-\omega_{d}$, $\Delta_{a}=\omega_{a}-\omega_{d}$,
and $g(\varphi_{1})$ given by either Eq. (\ref{eq:gd}) or Eq. (\ref{eq:gc}) depending on the coupling mechanism. The environmental noise can play an important role in the stationary state of the coupled system. We treat the noise by the Lindblad form $\kappa\mathcal{L}(\hat{a})\rho+\gamma_{d}\mathcal{L}(\sigma_{-})\rho+\gamma_{p}\mathcal{L}(\sigma_{z})\rho$ in the master equation. which includes dissipation of the junction resonator with the rate $\kappa$, decay of the TLS with the rate $\gamma_{d}$, and dephasing of the TLS with the rate $\gamma_{p}$.

The model discussed above describes a typical cavity QED system \cite{RaimondRMP2001,HoodScience2000}. The junction resonator mode acts as a high-Q cavity driven by microwave source and the TLS acts as a two-level atom coupled to the resonator mode with a JC-type of interaction. Because of the coupling, the transmission through the junction resonator is imprinted by the properties of the TLS. As we will show below, measurement of the microwave transmission provides an effective probe, or a microscope,  for the TLS.  Note that  the frequencies of the TLSs were observed to be separated by $200\,\textrm{MHz}$ on average \cite{MartinisPRL2005}. With a coupling magnitude of $g \sim 10\,\textrm{MHz}$, it is reasonable to assume that only one TLS satisfies the near-resonance condition: $|\omega_{a}-\omega_{c} |\sim g $  and affects the transmission in the junction resonator significantly.  The TLSs that are far off-resonance from the junction resonator mode  induce small ac-Stark shifts on the order of $g^{2}/|\omega_{a}-\omega_{c}|$.

\emph{Microwave transmission in the junction resonator.} To quantitatively illustrate the effect of the TLS coupling on the  transmission in the junction resonator,  we study the above system in the  ``bad cavity'' limit with $\gamma_{d},\gamma_{p}\ll\kappa$, i.e. the dissipation of the
junction resonator is much faster than that of the TLS \cite{WangVyasPRA1995}.  As a result, the junction resonator mode adiabatically follows the dynamics of the TLS which is governed by the Bloch equation:\begin{equation}
\frac{d\langle\vec{\sigma}\rangle}{dt}=A_{2}(\langle\vec{\sigma}\rangle+\vec{B})\label{eq:bloch}\end{equation} with the Pauli matrices $\vec{\sigma}=(\sigma_{z},\sigma_{+},\sigma_{-})^{T}$.  The dynamic matrix is \[
A_{2}=\left(\begin{array}{ccc}
-\gamma_{1} & -i\Omega_{r}^{*} & i\Omega_{r}\\
-\frac{i}{2}\Omega_{r} & i\Delta-\gamma_{2} & 0\\
\frac{i}{2}\Omega_{r}^{*} & 0 & -i\Delta-\gamma_{2}\end{array}\right)\]
and the offset vector is $\vec{B}=A_{2}^{-1}\cdot(-\gamma_{1},0,0)^{T}$ with \begin{eqnarray}
\Omega_{r} & = & \frac{i2g\epsilon}{\kappa-i\Delta_{c}}\label{eq:Or}\\
\Delta & = & \Delta_{a}-\frac{g^{2}\Delta_{c}}{\kappa^2+\Delta_{c}^2} \label{eq:Delta} \\ 
\gamma_2 & = & \frac{\gamma_{d}}{2}+\gamma_{p}+\frac{g^{2}\kappa}{\kappa^2+\Delta_{c}^2}\label{eq:gamma2}\end{eqnarray}
and $\gamma_{1} = 2(\gamma_{2}-\gamma_{p}) $ as the dressed parameters in the Bloch equation. The coupling between the TLS and the junction resonator mode modifies the detuning $\Delta$ and the decoherence $\gamma_{1,2}$ of the TLS, and results in an effective driving $\Omega_r$ on the TLS that depends linearly on the driving magnitude. The stationary state of the TLS  is $\langle\vec{\sigma}\rangle_{ss}=-\vec{B}$.  Defining $\eta=\Omega_{r}/(\Delta+i\gamma_{2})$,  we have  $B_{z}\approx1-|\eta|^{2}\gamma_{2}/\gamma_{1}$ and $B_{+}=B_{-}^{\star}\approx\eta/2$ under the weak driving condition $|\eta|\ll1$.

\begin{figure}
\includegraphics[%
  bb=72bp 108bp 576bp 450bp,
  clip,
  width=7.5cm]{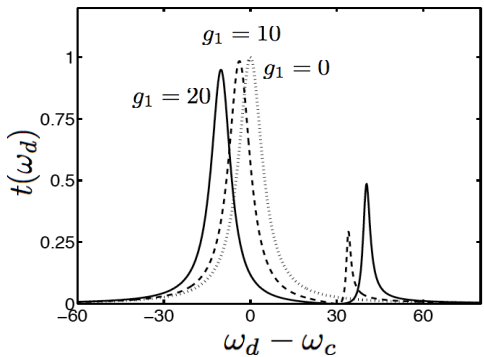}
\caption{Transmission in the junction resonator with the coupling strength $g_{d}=0$ (dotted curve), $g_{d}=10$ (dashed curve), and $g_{d}=20$ (solid curve) respectively versus the driving frequency $\omega_d$.  The parameters are $\omega_{c}=7\,\textrm{GHz}$, $\Delta_{ac}=30\,\textrm{MHz}$, $\gamma_{d}=0.2\,\textrm{MHz}$, $\gamma_{p}=0.1\,\textrm{MHz}$, $\kappa=5\,\textrm{MHz}$, and $\epsilon=1\,\textrm{MHz}$ for a TLS at the position $r_{d}=0.2$. }\label{fig2}
\end{figure}
Below, we derive  the transmission $t(\omega_d)=|\kappa\langle\hat{a}\rangle_{ss}/\epsilon|^{2}$ at the driving frequency $\omega_d$, depending on the ratio of the output field  $\langle\alpha\rangle_{ss}=(\epsilon+g\langle\sigma_{-}\rangle_{ss})/(\kappa+i\Delta_c)$  and the driving field. In Fig. \ref{fig2}, we plot the transmission versus the driving frequency at various values of applied magnetic flux  for the coupling magnitude $g(\varphi_1)=g_{d}$ defined in Eq.~(\ref{eq:gd}). Let  $\omega_p^{1,2}$ be the position of the resonance peaks in the transmission. At $g_d=0$ with no coupling, a resonance peak appears at $\omega_d=\omega_c$.  At  $\varphi_{1}=\pi/2$, the coupling increases to $g_{d}=10\,\textrm{MHz}$ and two resonance peaks appear with the separation $\omega_{p}^{2}-\omega_{p}^{1}=38\,\textrm{MHz}$. At  $\varphi_{1}=\pi$, the coupling is  $g_{d}=20\,\textrm{MHz}$ and the separation of the two resonance peaks becomes $\omega_{p}^{2}-\omega_{p}^{1}=50\,\textrm{MHz}$. To explain this result, we make  the following approximation,
\begin{equation}
t(\omega_{d})\approx \frac{\kappa^{2}\Delta_{a}^{2}}{(\Delta_{a}\Delta_{c}-g^{2})^{2}+\kappa^{2}\Delta_{a}^{2}}.\label{eq:tsimp}\end{equation} 
under the condition $\gamma_{d},\,\gamma_{p}\ll g_{1},\,\kappa,\,|\Delta_{a}|,\,|\Delta_{c}|$.  With no coupling at $g_d=0$, $t(\omega_{d})=\kappa^{2}/(\kappa^{2}+\Delta_{c}^{2})$ with  one resonance peak at $\omega_{d}=\omega_{c}$.  With finite coupling, the approximated transmission in Eq.~(\ref{eq:tsimp}) has two resonance peaks with frequencies satisfying $\Delta_{a}\Delta_{c}-g^{2}=0$., i.e.  $\omega_{p}^{(1,2)}=(\omega_{a}+\omega_{c}\pm\sqrt{\Delta_{ac}^{2}+4g^{2}})/2$ respectively, and $\Delta_{ac}=\omega_{a}-\omega_{c}$. This shows very good agreement with the plots in Fig.~\ref{fig2}. For the  coupling given by Eq.(\ref{eq:gc}), the coupling and hence the transmission are not affected by the applied magnetic flux.  Note that in Eq.~(\ref{eq:tsimp}) the transmission reaches a minimum at $\omega_d=\omega_{a}$ with $t(\omega_{a})$ approaches zero.  The position of this minimum can be used to determine the frequency of the TLS.

\emph{Noise spectrum  of the junction transmission.} The time correlation function of the  junction transmission can be measured to study the noise spectrum of the transmission, and subsequently the properties of the TLS. Let $X(t)=(\hat{a}+\hat{a}^{\dagger})(t)$ be the linearized operator of the phase variable $(\delta+\varphi_1/2)$.  The symmetrized noise spectrum for the operator $X(t)$ can be written as $F_{XX}(\omega)=\int_{-\infty}^{\infty}dte^{i\omega t}\frac{1}{2}\langle X(t)X+XX(t)\rangle-\langle X(t)\rangle\langle X\rangle $.  In the ``bad cavity'' limit \cite{WangVyasPRA1995}, we have,\begin{equation}
\frac{d}{dt}\hat{a}_{\alpha}(t)=-(\kappa-i\alpha\Delta_{c})\hat{a}_{\alpha}(t)+\epsilon+g\sigma_{\alpha}(t)\label{eq:transform}\end{equation}
with $\alpha\in\{-,+\}$, $\hat{a}_{-}=\hat{a}$ and $\hat{a}_{+}=\hat{a}^{\dagger}$, which forms a linear transformation between the Pauli matrices of the TLS and the operators of the junction resonator.  The time correlation function $\langle X(t)X\rangle $ (and $\langle X X(t)\rangle$) can then be derived from the time correlation functions of the Pauli matrices of the TLS.

Following the quantum regression theorem \cite{LaxPhysRev1963}, the time correlation functions of the operators of the TLS can be derived from Eq.~(\ref{eq:bloch}) and the matrix $A_2$.   At weak driving $|\eta|\ll1$, the matrix $A_{2}$ can be approximated to be a diagonal matrix with the elements $\{A_{zz}, A_{++}, A_{--}\}$ (not shown) under a second order perturbation of $\eta$ ($\eta^\star$).  The time correlation functions can be then derived as  
\begin{equation}\langle\sigma_{\alpha}(t)\sigma_{\beta}\rangle=e^{A_{\alpha\alpha}t}\left(\langle\sigma_{\alpha}\sigma_{\beta}\rangle-B_{\beta}B_{\alpha}\right)+B_{\beta}B_{\alpha}\label{eq:QRT}\end{equation} 
and similarly for $\langle\sigma_{\alpha}\sigma_{\beta}(t)\rangle$. Among such time correlation functions,  the dominant contribution to $F_{XX}(\omega)$ is by $\langle\sigma_{-}\sigma_{+}\rangle\approx 1$. 

\begin{figure}
\includegraphics[  bb=90bp 180bp 504bp 480bp,
  clip,  width=7.5cm]{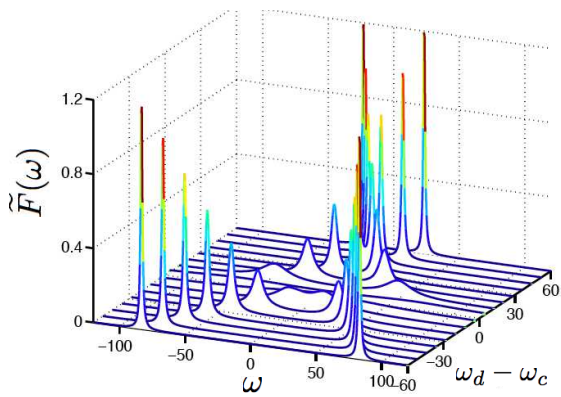}
\caption{Normalized noise spectrum  $\widetilde{F}_{\sigma_{-}\sigma_{+}}(\omega)$
at $g=20\,\textrm{MHz}$ and $|eta| \ll 1$ plotted for various driving frequencies. For parameters, see the caption of  Fig.~\ref{fig2}. }\label{fig3}
\end{figure}

We consider the normalized noise spectrum: $\widetilde{F}(\omega)=F_{XX}(\omega)(\kappa^{2}+(\omega-\Delta_{c})^{2})/g^{2}$.  It can be calculated that 
\begin{equation}
\widetilde{F}(\omega)\approx\frac{\gamma_{2}}{(\omega-\Delta)^{2}+\gamma_{2}^{2}}+\frac{\gamma_{2}}{(\omega+\Delta)^{2}+\gamma_{2}^{2}}\label{Fbw}\end{equation}
to an accuracy of $\textrm{O}(|\eta|^{2})$, directly associated with the parameters of the TLS. The frequencies of the resonance peaks in the noise spectrum are given by $\omega_{np}^{(1,2)}=\pm \Delta$ as defined in Eq.~(\ref{eq:Delta}).  The width of the resonance peaks is given by $\gamma_2$ as defined in Eq.~(\ref{eq:gamma2}). By measuring the resonances in the noise spectrum $\widetilde{F}(\omega)$, detailed characterization of the  TLS can be achieved. For example, at the driving frequency $\omega_d=\omega_c$ ($\Delta_c=0$), the position of the  resonances peaks is given by $\omega_{np}^{(1,2)}=\pm \Delta_{ac}$,  revealing the frequency $\omega_a$ of the TLS. At $\Delta_c\ne 0$ but $\Delta_c\sim \kappa$,  the position of the resonance peaks strongly depends on the coupling strength $g$. In Fig. \ref{fig3}, we plot $\widetilde{F}(\omega)$ for $g=20\,\textrm{MHz}$ over a range of driving frequencies.  When reducing the coupling strength to $g\ll \kappa$, the noise spectrum shows  two resonance peaks with frequencies linearly increasing with the driving frequency and crossing each other at $\omega_d=\omega_a$. 

\emph{Spectroscopic Measurements.} The spectrum of the TLS covers a broad range of frequency.  In order to clearly detect the variations in the coupling and the position of the TLSs, we need to control the resonant frequency of the junction resonator $\omega_c$ to sweep past the individual TLSs for each successive set of transmission measurements. This is achieved by varying the applied flux $\Phi_b$ in the RF-SQUID loop to adjust the total Josephson inductance in the RF-SQUID loop.   The effective frequency of the junction resonator can then be derived as \begin{equation} 
\omega_c=\sqrt{\frac{cos(2\pi*\Phi_b/\Phi_o)}{L_{J}C_{J}} +
\frac{1}{L_{b}C_{J}}}\label{eq:wc_aj}\end{equation}
where $L_{J}$ is  the intrinsic inductance of the Josephson junction when $\Phi_b=0$ and $L_{b}$ is the self-inductance of the RF-SQUID bias loop. 

Once a TLS is chosen and the frequency of the TLS is close to the frequency of the junction resonator mode,  we want to keep the frequency $\omega_c$ from changing during the transmission measurements while continuously varying the magnetic field $\Phi_1$.  The bias flux $\Phi_b$ in the RF-SQUID loop again provides crucial control of the frequency $\omega_{c}$.   Adjusting the flux $\Phi_1$ inside the tunnel barrier, $\omega_{c}$ is shifted as we discussed previously. But by tuning the flux $\Phi_b$ in the RF-SQUID loop, this frequency shift can be compensated to stay in a constant energy contour. 

\emph{Conclusions.} We present a cavity QED scheme for studying the properties of the TLSs in the tunnel barrier of a Josephson junction. The high-Q oscillator mode of the junction resonator acts as a microscope for probing the spectral, spatial, and coupling properties of the TLSs by the measured the microwave transmission in the junction.
In particular, our study shows that the coupling mechanism between the junction and the TLS can be determined by applying a magnetic field in the plane of the junction
tunnel barrier.

L T.  is supported by Karel Urbanek Fellowship of Stanford Univ., SORST program of Japan Science of Technology Corporation (JST), and Special Coordination Funds for Promoting Science and Technology of Univ. Tokyo.

\end{document}